\documentclass[twocolumn,aps,prc,superscriptaddress,nofootinbib,floatfix]{revtex4-2}

\usepackage[T1]{fontenc}
\usepackage[utf8]{inputenc}
\usepackage{lmodern}
\usepackage{amsmath,amssymb,bm}
\usepackage{graphicx}
\usepackage{xcolor}
\usepackage[colorlinks=true,linkcolor=blue,citecolor=blue,urlcolor=blue]{hyperref}
\usepackage{braket}
\usepackage{microtype}

\graphicspath{{figs/}{./}}

\newcommand{\avg}[1]{\left\langle #1 \right\rangle}

\newcommand{\figplaceholder}[2][0.92\linewidth]{%
  \fbox{\parbox[c][0.17\textheight][c]{#1}{\centering\small Placeholder for figure\\\texttt{\detokenize{#2}}}}%
}
\newcommand{\maybeincludegraphics}[2][]{%
  \IfFileExists{#2}{\includegraphics[#1]{#2}}{\figplaceholder{#2}}%
}

\usepackage{tikz,xcolor,hyperref}

\definecolor{lime}{HTML}{A6CE39}
\DeclareRobustCommand{\orcidicon}{
	\begin{tikzpicture}
	\draw[lime, fill=lime] (0,0) 
	circle [radius=0.16] 
	node[white] {{\fontfamily{qag}\selectfont \tiny ID}};
	\draw[white, fill=white] (-0.0625,0.095) 
	circle [radius=0.007];
	\end{tikzpicture}
	\hspace{-2mm}
}
\foreach \x in {A, ..., Z}{%
	\expandafter\xdef\csname orcid\x\endcsname{\noexpand\href{https://orcid.org/\csname orcidauthor\x\endcsname}{\noexpand\orcidicon}}
}
\foreach \x in {A, ..., Z}{%
	\expandafter\xdef\csname orcid\x\endcsname{\noexpand\href{https://orcid.org/\csname orcidauthor\x\endcsname}{\noexpand\orcidicon}}
}

\begin{document}

% \title{Hadron polarization in the FAIR/RHIC-BES energy region and the effect of the EoS}
\title{Hadron polarization and equation of state at FAIR/RHIC-BES energies}

\author{Dai-Neng Liu\orcidA{}}
\affiliation{Key Laboratory of Nuclear Physics and Ion-beam Application (MOE), Institute of Modern Physics, Fudan University, Shanghai 200433, China}
\affiliation{Shanghai Research Center for Theoretical Nuclear Physics, NSFC and Fudan University, Shanghai 200438, China}
\affiliation{Institut f\"ur Theoretische Physik, Goethe-Universit\"at Frankfurt, Max-von-Laue-Str.~1, D-60438 Frankfurt am Main, Germany}
\affiliation{Helmholtz Research Academy Hesse for FAIR (HFHF), GSI Helmholtzzentrum f\"ur Schwerionenforschung GmbH, Campus Frankfurt, Max-von-Laue-Str.~12, 60438 Frankfurt am Main, Germany}

\author{Jan Steinheimer\orcidB{}}
\email{j.steinheimer-froschauer@gsi.de}
\affiliation{GSI Helmholtzzentrum f\"ur Schwerionenforschung GmbH, Planckstr. 1, D-64291 Darmstadt, Germany}
\affiliation{Frankfurt Institute for Advanced Studies, Ruth-Moufang-Str. 1,  60438 Frankfurt am Main, Germany}

\author{Kai-Jia Sun\orcidC{}}
\email{kjsun@fudan.edu.cn}
\affiliation{Key Laboratory of Nuclear Physics and Ion-beam Application (MOE), Institute of Modern Physics, Fudan University, Shanghai 200433, China}
\affiliation{Shanghai Research Center for Theoretical Nuclear Physics, NSFC and Fudan University, Shanghai 200438, China}

\author{Jin-Hui Chen\orcidD{}}
\email{chenjinhui@fudan.edu.cn}
\affiliation{Key Laboratory of Nuclear Physics and Ion-beam Application (MOE), Institute of Modern Physics, Fudan University, Shanghai 200433, China}
\affiliation{Shanghai Research Center for Theoretical Nuclear Physics, NSFC and Fudan University, Shanghai 200438, China}

\author{Yu-Gang Ma\orcidE{}}
\email{mayugang@fudan.edu.cn}
\affiliation{Key Laboratory of Nuclear Physics and Ion-beam Application (MOE), Institute of Modern Physics, Fudan University, Shanghai 200433, China}
\affiliation{Shanghai Research Center for Theoretical Nuclear Physics, NSFC and Fudan University, Shanghai 200438, China}
\affiliation{School of Physics, East China Normal University, Shanghai 200241, China}

\author{Marcus Bleicher\orcidF{}}
\email{bleicher@itp.uni-frankfurt.de}
\affiliation{Institut f\"ur Theoretische Physik, Goethe-Universit\"at Frankfurt, Max-von-Laue-Str.~1, D-60438 Frankfurt am Main, Germany}
\affiliation{Helmholtz Research Academy Hesse for FAIR (HFHF), GSI Helmholtzzentrum f\"ur Schwerionenforschung GmbH, Campus Frankfurt, Max-von-Laue-Str.~12, 60438 Frankfurt am Main, Germany}

\date{\today}

\begin{abstract}
The $\Lambda$ global polarization indicates that hot and dense matter created in non-central heavy-ion collisions carries large orbital angular momentum. 
However, the relation between hadronic polarization and the medium's collective rotation remains to be validated. 
Using the UrQMD transport model, we calculate the thermal vorticity-induced polarization of $\Lambda$s in Ag+Ag and Au+Au collisions from $\sqrt{s_{\rm NN}}=2.24$-$7.7$ GeV and a range of centralities. Two different equations of state used in the UrQMD simulation are compared: one resembles a hadron resonance gas, while the other is based on the chiral mean field (CMF) model, providing a more realistic description of dense nuclear matter including a chiral transition that is consistent with lattice QCD expectations. The polarization is sensitive to the equation of state and a softer EoS leads to smaller values. In addition, we show that the $\Lambda$ polarization in the experimental acceptance and centrality selection does not decrease for even lower beam energies. Our results indicate that the process leading to the large vorticity is a result of the large shear in the baryon current created by its stopping. 

\end{abstract}

\maketitle

\section{Introduction}

The observation of global $\Lambda$ polarization in heavy-ion collisions leads to the conclusion that the matter created in non-central nuclear reactions is the most vortical fluid observed in the laboratory~\cite{STAR:2017ckg,STAR:2018gyt}. Moreover, further measurement of the polarization of $\Xi$ and $\Omega$ hyperons confirmed the existence of a local vorticity-based polarization picture~\cite{STAR:2020xbm,Li:2021zwq}. Since the spin of emitted hadrons couples to the local thermal vorticity, polarization measurements have become a central tool for studying the dynamics of hot and dense QCD matter~\cite{Liang:2004ph,Liang:2004xn,Becattini:2013fla,Becattini:2013vja,Csernai:2014ywa,Xie:2016fjj,Becattini:2016gvu,STAR:2019erd,ALICE:2019aid,Weickgenannt:2020aaf,STAR:2022fan,Becattini:2024uha,Niida:2024ntm,Chen:2024afy,Huang:2024ffg,Sun:2025oib,STAR:2025njp,Liang:NST,Zheng:PRC,Chen:2026gka,Liu:2025kpp}.
On the other hand, several aspects of hadron polarization are still under investigation. This includes the observed large spin alignment of the vector-meson $\phi$ \cite{Liang:2004xn,STAR:2022fan} and the question of how and why spin would be locally equilibrated over such a wide range of beam energies \cite{Florkowski:2018fap,Becattini:2021iol,Liu:PLB}. In addition, the question of a baseline remains relevant, i.e., can any polarization also be observed in elementary p+p reactions, where transverse hyperon polarization has long been known and is now being revisited \cite{Bunce:1976yb,STAR:2025jwc,Liu:2026rye}. Further, it is often assumed that the thermal vorticity may vanish as the beam energy of the nuclear collisions decreases, which would lead to a maximum in the $\Lambda$ polarization with beam energy. However, studies at beam energies of a few MeV have suggested the formation of rotating compound nuclei \cite{Janzen:1994zz,Ciemala:2015xua} and this begs the question whether a maximum exists and where it is located. Recent measurements of hyperon polarization at SIS18 energies \cite{Deng:2021miw,Guo:2021udq,HADES:2022enx} have shown an even larger polarization than at the higher beam energies. To understand the origin of the vorticity and also its energy dependence down to the lowest beam energies would therefore be invaluable for the interpretation of it at the highest beam energies.

To study and understand the remaining open questions, dynamical model descriptions of heavy ion collisions are necessary. In the recent simulations, the local vorticity field is either obtained from  hydrodynamic~\cite{Csernai:2014ywa,Karpenko:2016jyx,Xie:2017upb,Fu:2020oxj,Ivanov:2025izv} or microscopic transport model~\cite{Jiang:2016woz,Li:2017slc,Wei:2018zfb,Shi:2017wpk,Vitiuk:2019rfv,Deng:2020ygd,Huang:2020xyr,Guo:2021udq,Deng:2021miw,Yi:2026rbz} calculations, leading to a resulting polarization comparable with data. 
One open problem is how the collective vorticity of the emitting source is transferred to the hadronic polarization. In the above models, this is essentially done via the assumption of local equilibration, including spin. An argument that is often put forward in this context is that this local equilibration of the spin may be related to hadronization and thus the polarization measurement may be a probe for deconfinement.
This may be questionable when going to very low beam energies.

One natural candidate to understand how vorticity can be formed at low beam energies is the reaction-plane tilt of the freeze-out source, caused by the asymmetric baryon stopping. Especially, azimuthally sensitive femtoscopy has been shown to be able to determine the orientation of the emission ellipsoid relative to the beam axis~\cite{Lisa:2000ip,Lisa:2005dd,Khyzhniak:2024chj}. The tilted freeze-out geometry has long been associated with the dynamics responsible for directed flow and with the stopping-driven shear between projectile and target matter~\cite{Bozek:2010bi,Retiere:2003kf}. 
The twisted shape of the freeze-out source has been studied via pion–pion femtoscopic correlations~\cite{Lisa:2011na,Mount:2010ey,Graef:2013wta} and charmonium directed flow~\cite{Chen:2019qzx}. In addition, the vorticity in heavy-ion collisions has been investigated independently through proton–pion correlations~\cite{Savchuk:2025kuk}.

In a recent work, it was argued that through the use of the thermal vorticity, the measured polarization can be sensitive to the QCD equation of state (its degrees of freedom) and that only a pure hadron gas equation of state (i.e. one without a phase transition or crossover) is able to describe the measured $\Lambda$ polarization \cite{Yi:2026rbz}.

In this work, we will address the above two challenges and use the UrQMD model~\cite{Bass:1998ca,Bleicher:1999xi}, which has been extended to incorporate a realistic chiral mean field (CMF) equation of state to calculate the polarization of $\Lambda$ hyperons from RHIC down to the very lowest SIS18 beam energies. We will show that the sensitivity to the EoS has to be studied in a way that ensures consistency between the EoS and the corresponding temperature as a function of energy density. Using this approach, we also make predictions for Au+Au collisions at ${E_{lab}=0.8A}$~GeV to explore the low energy asymptotic behavior of the polarization.  

\section{Method and numerical setup}

\subsection{Vorticity-induced polarization}

We simulate Au+Au collisions with the UrQMD transport model~\cite{Bass:1998ca,Bleicher:1999xi} and follow previous works to calculate the polarization from the thermal vorticity using transport models~\cite{Li:2017slc,Li:2021zwq,Huang:2020xyr,Deng:2021miw,Yi:2026rbz,Deng:arxiv,Xi:2023isk}. The calculations are performed using a coarse-graining prescription where we use a four-dimensional lattice with spatial lattice spacing $\Delta x = \Delta y = \Delta z = 1~\mathrm{fm}$ in the range (-50, 50) fm and temporal extent $ \Delta t = 1~\mathrm{fm}/c$ up to 50 fm/$c$.
In each space-time cell, the four-vector flow velocity $u^\mu$  and temperature $T$ of the local rest-frame are extracted from the event-averaged energy density, net-baryon density, and momentum density~\cite{Rischke:1995ir}, using the equation of state from the CMF model~\cite{Steinheimer:2024eha,Steinheimer:2025hsr,Negreiros:2026ode}. It is important to note that the equation of state that is used to find the local temperature and flow field is the same as that used for the interactions in the UrQMD model to maintain consistency.

The polarization of $\Lambda$ hyperon can then be calculated from the thermal vorticity at the space-time point where the $\Lambda$ is emitted, i.e.\ the point of its last scattering, which is taken from the UrQMD simulations. Once these points are defined, the kinetic and thermal vorticity tensors can be calculated.

For the relativistic case, the kinetic vorticity tensor is defined as
\begin{equation}
\omega_{\mu\nu}=-\frac{1}{2}\left(\partial_\mu u_\nu-\partial_\nu u_\mu\right),
\label{eq:kinvort}
\end{equation}
and the thermal vorticity tensor as
\begin{equation}
\varpi_{\mu\nu}=-\frac{1}{2}\left(\partial_\mu \beta_\nu-\partial_\nu \beta_\mu\right),
\label{eq:thermvort}
\end{equation}
where $\beta_\mu \equiv u_\mu/T$ with $T$ being temperature.
For spin-$1/2$ hadrons at local thermal equilibrium and to first order in gradients, the mean spin vector at the emission point $x$ and momentum $p$ is~\cite{Becattini:2013fla,Becattini:2016gvu}
\begin{equation}
S^\mu(x,p) = -\frac{1}{8m} (1-n_F)\,\epsilon^{\mu\nu\rho\sigma}p_\nu\,\varpi_{\rho\sigma}(x),
\label{eq:spinvector}
\end{equation}
where $n_F$ is the Fermi--Dirac distribution and has the approximation $1-n_F\approx1$ \cite{Becattini:2013vja}. In the present beam-energy range, the global observable is dominated by the reaction-plane component associated with $\varpi_{zx}$ which quantifies the gradient of the longitudinal velocity along the $x$-axis, i.e.\ longitudinal shear caused by baryon stopping.

To compare with the experimentally measured hyperon polarization, we transform the spatial spin vector to the particle's rest frame from c.m.~frame as,
\begin{equation}
\mathbf{S}^{*} = \mathbf{S} - \frac{\mathbf{p}\cdot\mathbf{S}}{E_p(E_p+m)}\,\mathbf{p},
\label{eq:restframe}
\end{equation}
where \(E_p\), \(\mathbf{p}\), and \(m\) denote the energy, momentum, and mass of the particle, respectively,
and define the global polarization for spin-$1/2$ hadrons as
\begin{equation}
\mathcal{P}_H = 2\,\avg{\mathbf{S}^{\,*}\cdot \hat{\mathbf{J}}},
\label{eq:globalP}
\end{equation}
where \(\hat{\mathbf{J}}=\mathbf{J}/|\mathbf{J}|\) denotes the unit vector along global angular-momentum direction, and $H$ denotes the hadron species, primarily $\Lambda$ in the present work. As experiments are usually not capable of distinguishing between direct $\Lambda$s and those coming from the decay of $\Sigma^0$-hyperons, we included all decayed $\Lambda$ from $\Sigma^0$ in UrQMD analysis as well. One should note that the $\Sigma^0$ decay may modify the observed polarization of the $\Lambda$ \cite{Becattini:2016gvu,Karpenko:2016jyx,Xia:2019fjf,Becattini:2019ntv}. In the present work, we will neglect all possible feed-down effects and the microscopic hadronic scatterings \cite{Sung:2024vyc}, both of which lead to polarization modifications.

\subsection{The equations of state}
The equation of state that is used for the calculation of the local temperature and flow fields is the same that was recently implemented in the QMD part of the UrQMD transport model \cite{Steinheimer:2025hsr}. This ensures that the dynamics is consistent with the extracted temperatures. This equation of state, which we will refer to as the CMF EoS, is based on a chiral mean field model where chiral symmetry restoration is realized through a parity doublet description of baryons \cite{Steinheimer:2011ea}. For a more detailed description of the model, its implementation into UrQMD and the used parameter sets, as well as the resulting phase diagram, we refer to \cite{Steinheimer:2025hsr,Steinheimer:2024eha}. It is important to note that the CMF EoS not only provides a good description of nuclear matter properties, important to describe flow data and particle multiplicities in heavy-ion collisions in the SIS/FAIR energy range, its extension to finite temperatures is also consistent with the thermodynamics of lattice QCD at vanishing net baryon density.

To obtain the flow velocity and temperature in the computational frame we use a tabulated form of the CMF EoS (pressure and temperature) in terms of energy density and net baryon density \footnote{Note, that such a table is also used in the hydrodynamic evolution of the UrQMD-hybrid model.}. 
Alternatively, a hadron resonance gas (dubbed HRG EoS or CASCADE mode) is employed, where all interaction potentials are switched off.

To calculate the flow velocity, we assume that the energy-momentum flow is the same as the baryon flow and calculate the pressure and temperature in the Landau frame. In that case, the energy density,  net-baryon density, and momentum density ($\epsilon^{lab}$,  $n^{lab}_B$, and $\mathbf{k}$) are related to the local rest frame energy density $\epsilon$, net-baryon density $n_B$, and fluid velocity $\mathbf{v}$ as 
\begin{equation}
    \begin{aligned}
    \mathbf{v}&=\frac{\mathbf{k}}{\epsilon^{lab}+p(\epsilon,n_B)},\\ 
    \epsilon&=\epsilon^{lab}-\mathbf{v}\cdot\mathbf{k},\\ 
    n_B&=n_B^{lab}/\gamma,
\end{aligned}
\end{equation}
where $\mathbf{v}$ represents the fluid velocity of a local fluid cell, $p$ is the pressure given by the equation of state, and $\gamma$ denotes the Lorentz factor. We solve the above equations for every point in our four-dimensional grid, obtaining the relevant quantities $u^\mu=\gamma(1,\mathbf{v})$ and $T$.

\section{Results}

\subsection{Vorticity}

\begin{figure}
    \centering
    \includegraphics[width=1\linewidth]{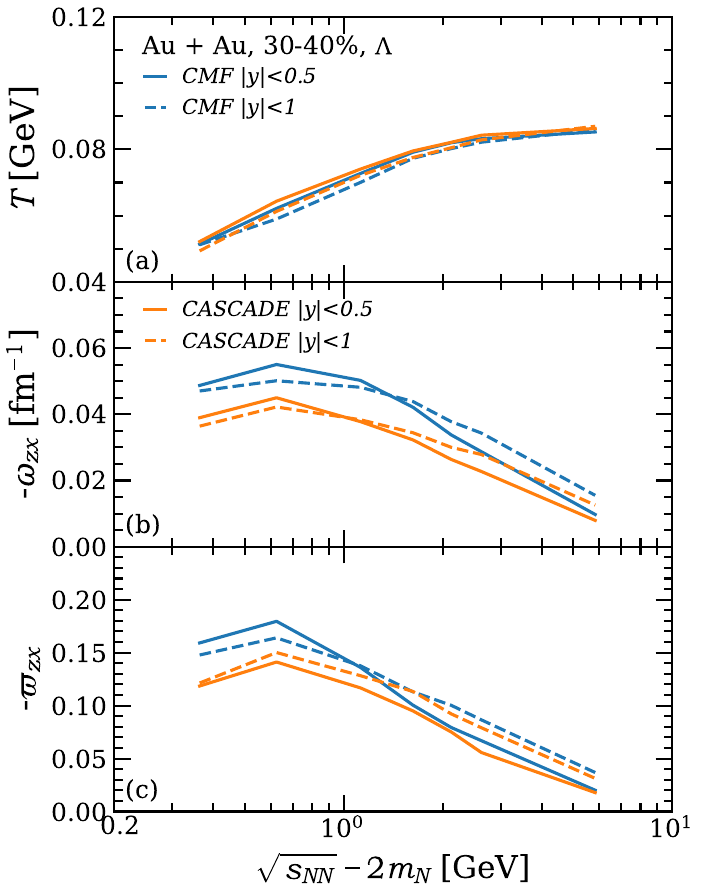}
    \caption{Comparison of UrQMD calculations with the CMF EoS and HRG/CASCADE EoS in two different rapidity windows as a function of collision energy. The values are obtained at the freeze-out space-time points of the $\Lambda$-hyperons. Upper panel: Average temperature,  middle panel: Average kinetic vorticity, 
    lower panel: Average thermal vorticity.}
    \label{fig:vort}
\end{figure}

First we exploit the different ingredients for polarization separately. Figure \ref{fig:vort} shows the average values of the temperature, vorticity, and thermal vorticity, at the space-time points of the $\Lambda$ emission, as a function of beam energy. All quantities are averaged over all freeze-out points of $\Lambda$'s from the UrQMD simulations.

The results are obtained from Au+Au collisions within 30-40\% centrality. The mean temperatures of $\Lambda$ hyperons at their freeze-out points as a function of the beam energy are shown in Fig.~\ref{fig:vort}(upper panel). The blue lines present results with the CMF-EoS UrQMD, while the orange lines represent the CASCADE mode of UrQMD (i.e.\ all interaction potentials are turned off and the effective EoS corresponds to that of a hadron resonance gas). The solid lines correspond to a smaller rapidity acceptance of $|y|<0.5$ and the dashed lines are for $|y|<1$. 
In all scenarios, the average temperature at freeze-out increases with the collision energy, from 50 MeV to 90 MeV within the energy range 2.24 - 7.7 GeV (this is consistent with previous findings for those beam energies \cite{STAR:2008med,STAR:2017sal,Chen:2024aom,Inghirami:2019muf}). The temperatures also do not show a significant dependence on which EoS is used for the UrQMD simulation which indicates that the freeze-out densities are approximately independent of the underlying EoS. 

The average value of kinetic vorticity ($\braket{\omega_{zx}}$) and thermal vorticity ($\braket{\varpi_{zx}}$) are shown in the middle and lower panels of Fig~\ref{fig:vort}. Since they are determined at the freeze-out space-time point of $\Lambda$ hyperons the results differ from the maximum vorticity which can be obtained as was shown in~\cite{Jiang:2016woz,Deng:2020ygd,Yi:2026rbz}.
The $\Lambda$ freeze-out vorticity peaks at approximately $\sqrt{s_{NN}}=2.5$ GeV in our calculations and does only drop moderately towards even lower collision energies, staying rather large even for $\sqrt{s_{NN}}=2.24$ GeV. 
In the vorticity, we observe a significant difference between the UrQMD-CMF case and the HRG/CASCADE version. This difference can be attributed to the different flow in both scenarios, where the UrQMD-CMF simulation creates more radial as well as directed flow due to larger pressure gradients. 

Within the coarse-grained fields, the dominant contribution to the vorticity comes from the shear term related to $\partial_x u_z$, while the corresponding thermal-vorticity pattern broadly follows the kinetic-vorticity energy dependence. 

We also observe a small dependence of the vorticities to the applied cut in rapidity. At higher beam energies the vorticity will increase for a larger rapidity window, while at low beam energies it will decrease. This is likely due to the fact that at lower beam energies, for rapidity windows of $|y|< 1$ the whole fireball and parts of the spectator fragments are inside the acceptance. At higher beam energies, as the rapidity window is increased, parts with larger longitudinal flow are included in the acceptance, increasing the vorticity.

\subsection{Global $\Lambda$ polarization}

Next, we can directly evaluate the resulting polarizations of the $\Lambda$ and compare our results with experimental data. We show the global polarization of $\Lambda$ hyperons in Au+Au collisions at low energies in the interval $\sqrt{s_{NN}}=2.24-7.7$ GeV by using the UrQMD model with the CMF EoS and in HRG/CASCADE mode.

The centrality classes, defined in terms of the collision impact parameter, are taken from~\cite{HADES:2017def} for Au+Au collisions (for Ag+Ag collisions they were provided via private communication) and are summarized in Table~\ref{tab:imp}. 

\begin{table}[b]
\begin{tabular}{|c|c|c|c|c|c|}
\hline
Centrality class & 0-10\% & 10-20\% & 20-30\% & 30-40\% & 40-50\% \\\hline\hline
$b$ [fm] (Au+Au) & 0-4.7    &4.7-6.6     &6.6-8.1     & 8.1-9.3     & 9.3-10.4    \\ \hline
$b$ [fm] (Ag+Ag) & 0-3.8    & 3.8-5.4     & 5.4-6.6     & 6.6-7.6     & 7.6-8.5  \\
\hline
\end{tabular}
\caption{Impact parameters used for the comparison with the experimental results. The centrality range 0-50\% is used for Au+Au and Ag+Ag collisions.}\label{tab:imp}
\end{table}

As has been shown in previous studies, the model provides a good comparison with experimental data for the basic hadron observables in relativistic heavy-ion collisions, such as particle yields, rapidity ($y$) and transverse-momentum ($p_T$) distributions, centrality-dependent multiplicities, and anisotropic flow.

\begin{figure}[t]
    \centering
    \includegraphics[width=0.49\textwidth]{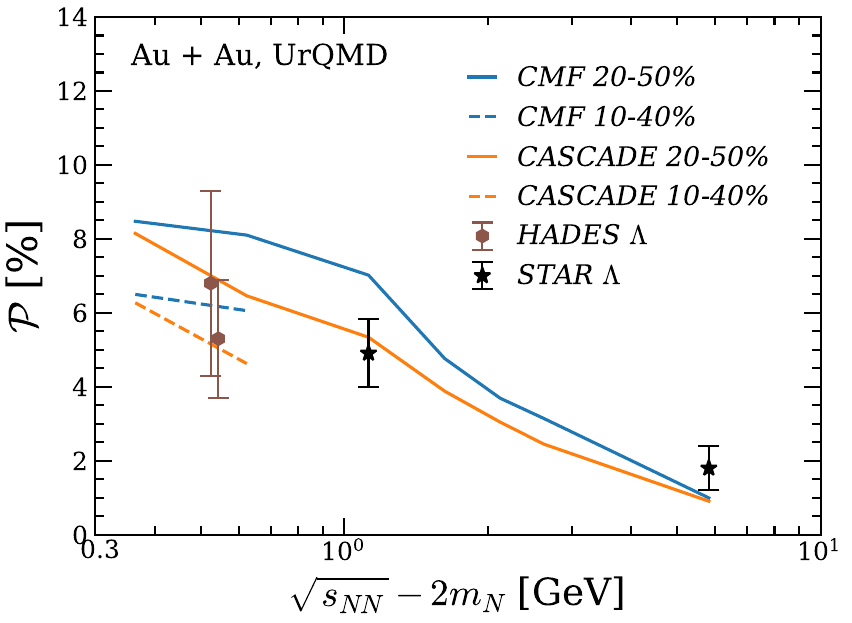}
    \caption{The global $\Lambda$ hyperon polarization $\mathcal{P}$ in noncentral Au+Au collisions as a function of the center-of-mass energy above threshold, $\sqrt{s_{NN}}-2m_{N}$. The solid (dashed) lines show UrQMD calculations in the CMF EoS (blue) and HRG/CASCADE EoS (orange) for the 20–50\% (10–40\%) centrality classes. The results are in reasonable agreement with the experimental data from the HADES (brown hexagons) and STAR (black stars) experiments.}
    % \caption{Energy dependence of the global polarization of $\Lambda$ hyperons in 20--50\% central Au+Au collisions. The simulation results decrease monotonically with $\sqrt{s_{\rm NN}}$ for both EoS (blue: CMF EoS and orange: HRG/CASCADE EoS). The results are in reasonable agreement with the experimental data from the HADES (10-40$\%$, dashed lines) and STAR (20-50\%, solid lines) experiments.}
    \label{fig:ener2050}
\end{figure}

Figure ~\ref{fig:ener2050} shows the energy dependence of $\Lambda$ hyperon polarization obtained from the thermal vorticity in Au+Au collisions for two different centrality selections, 20-50\% (solid lines, corresponding to STAR data) and 10-40$\%$ (dashed lines, corresponding to HADES data). For each case we always combine several sub-bins e.g. 20-30\%, 30-40\%, and 40-50\% to calculate the thermal vorticity and the spin vector of every $\Lambda$ hyperon, and finally take the average value of the polarization in the large centrality interval. For the comparison with STAR and HADES experimental data, the rapidity ranges and transverse momentum ranges are $y\in[-1, 1]$ and $p_\mathrm{T}\in[0.4, 3]$ GeV/$c$ for energies larger than 3 GeV~\cite{STAR:2017ckg}, $y\in[-0.2,1]$ and $p_\mathrm{T}>0.7$ GeV/$c$ for 3 GeV~\cite{STAR:2021beb}, and $y\in$[-0.5, 0.3] and $p_\mathrm{T}\in[0.2, 1.5]$ GeV/$c$ for the energy region smaller than 3 GeV~\cite{HADES:2022enx}. Note that HADES (2.24–2.5 GeV) reported results for the 10–40\% centrality range~\cite{HADES:2022enx}. The difference between these two centrality selections is shown as the gap between the solid and dashed lines.

The $\Lambda$ hyperon polarization increases with decreasing beam energy both with the CMF-EoS and HRG/CASCADE mode of UrQMD. The lowest collision energy 2.24 GeV in our calculation has the largest global polarization in the given centrality selection and we do not see any indication that the polarization would decrease at lower beam energies. Unfortunately at some point the energy available for hyperon production becomes so small that it would not be feasible to try and measure hyperon polarization.  
For all beam energies, the HRG/CASCADE calculation yields a smaller polarization than the one where the CMF EoS is used. This sensitivity suggests that polarization can help constrain the effective equation of state in the high-baryon-density regime. It also shows that the softer EoS (HRG/CASCADE) gives a smaller polarization than the harder CMF EoS. %which is in contrast to the statements in \cite{Yi:2026rbz}.

\begin{figure}[!t]
    \centering
    \includegraphics[width=0.48\textwidth]{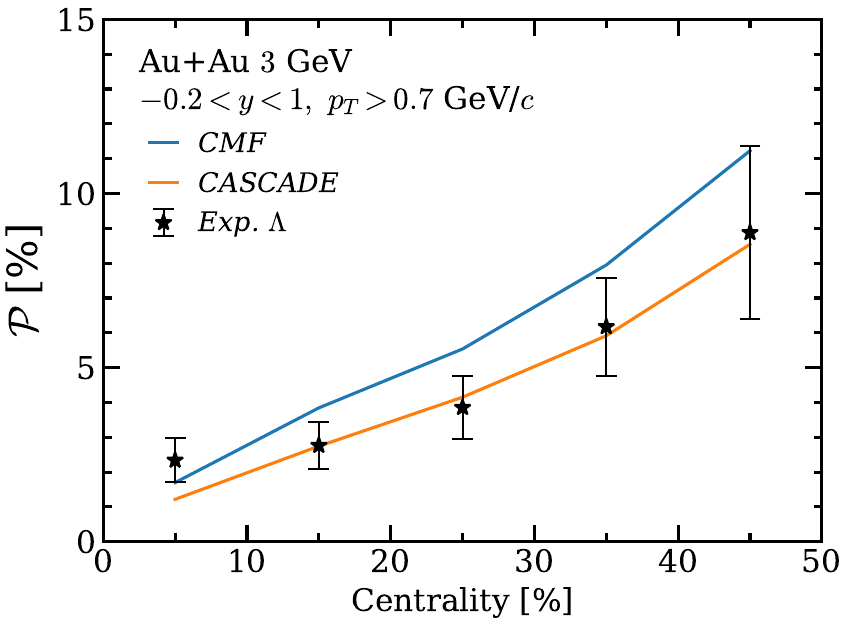}
    % \vspace{0.1em}
    \includegraphics[width=0.48\textwidth]{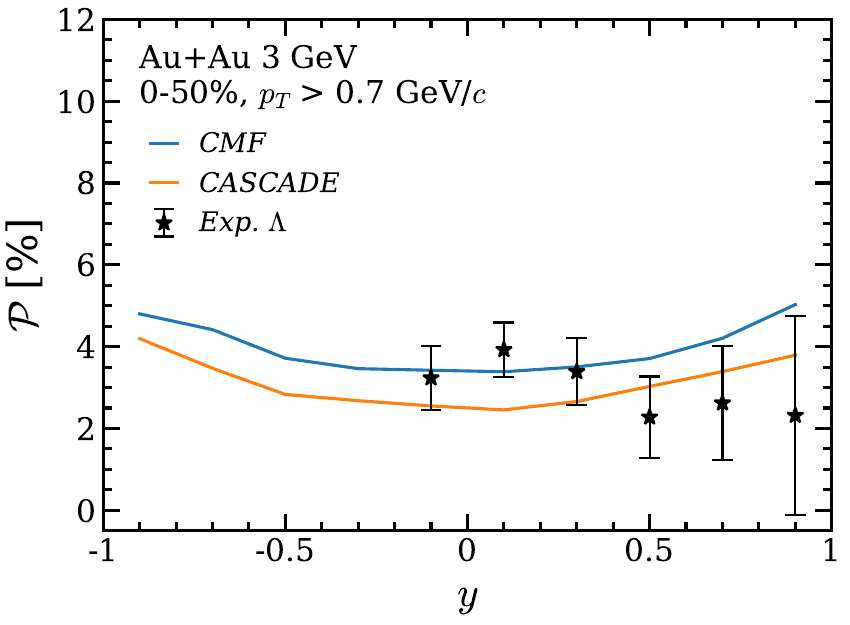}
    % \vspace{0.1em}
    \includegraphics[width=0.48\textwidth]{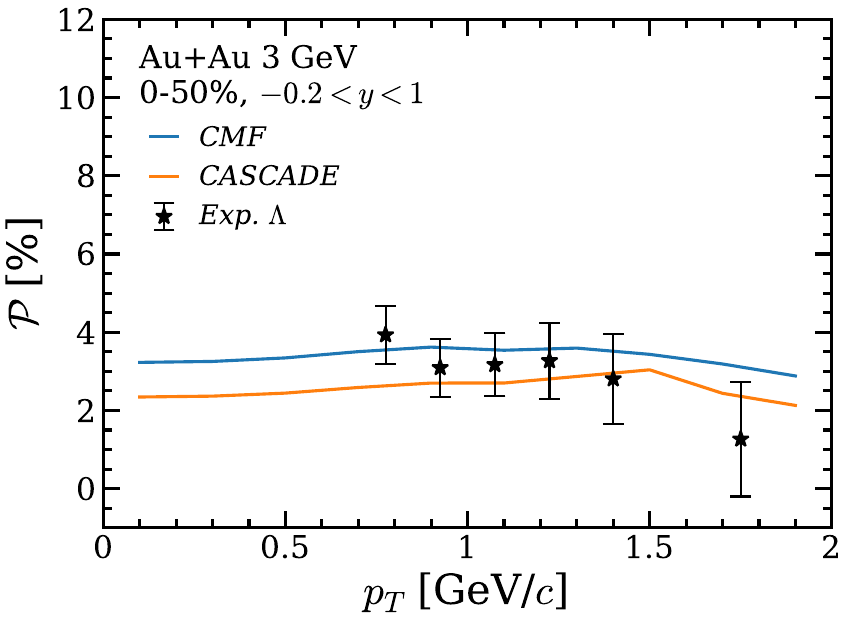}
    \caption{Centrality (upper panel), rapidity (middle panel), and transverse momentum (lower panel) dependence of $\Lambda$ polarization at $\sqrt{s_{\mathrm{NN}}}=3$ GeV in Au+Au collisions. $10^5$ events were used for each centrality bin. The experimental results are from \cite{STAR:2021beb}.}
    \label{fig:pt_y_cen}
\end{figure}

\begin{figure}[!t]
    \centering
    \includegraphics[width=0.48\textwidth]{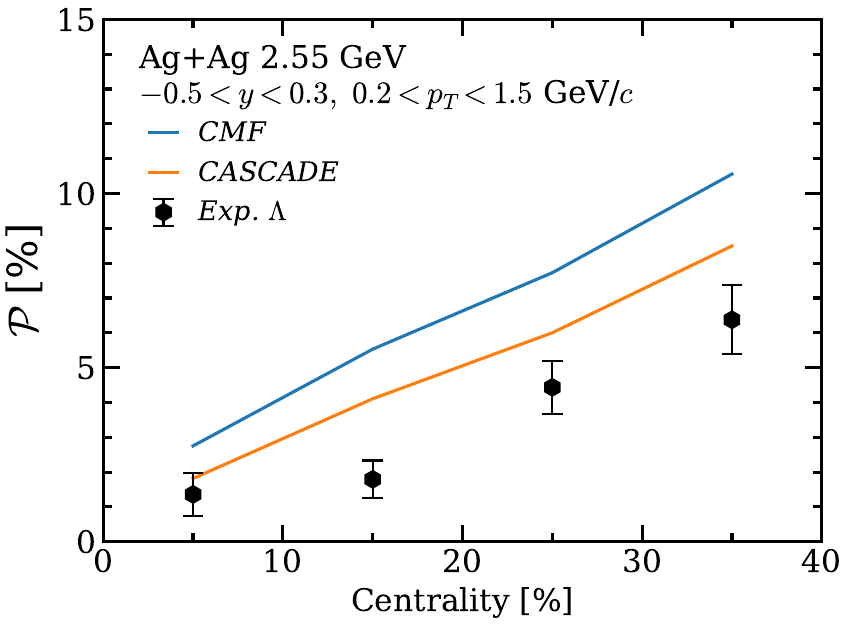}
    \includegraphics[width=0.48\textwidth]{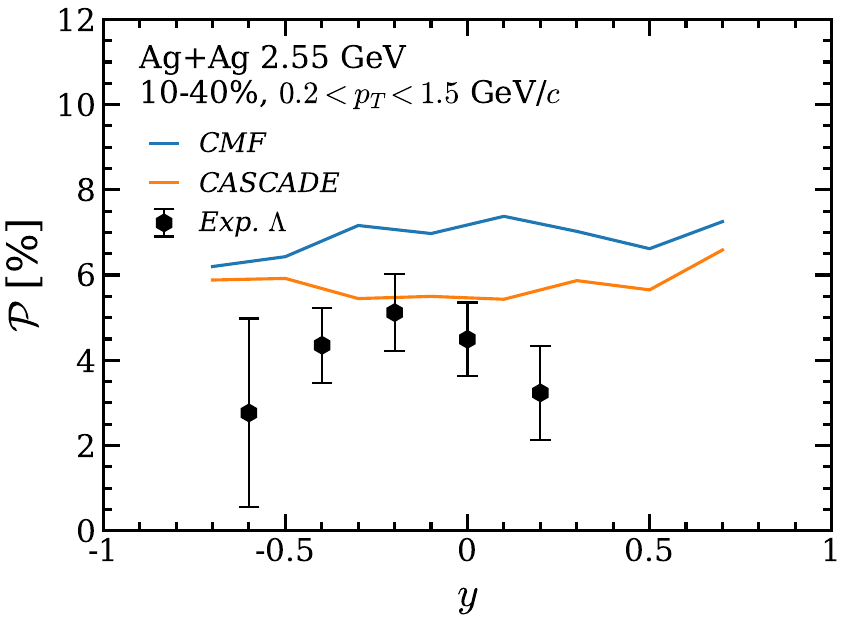}
    \includegraphics[width=0.48\textwidth]{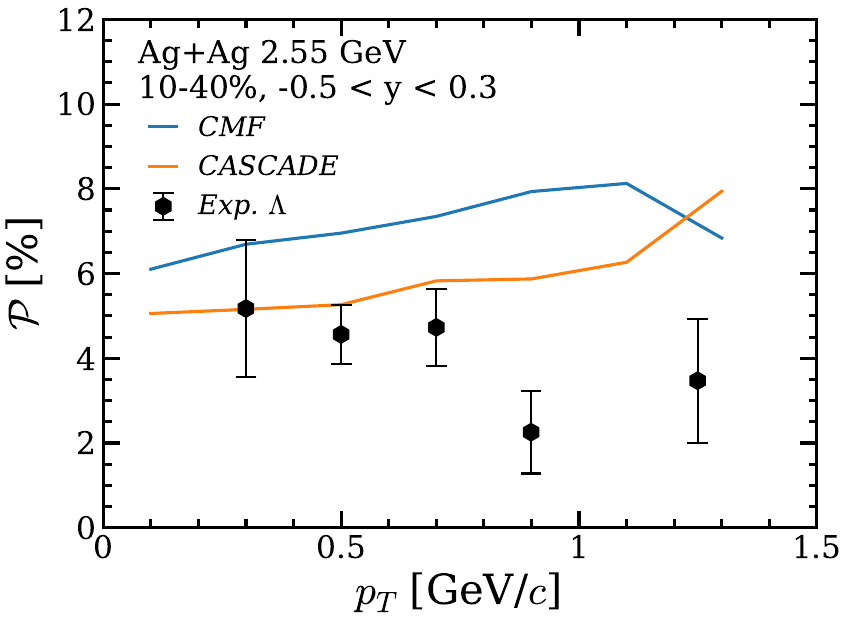}
    \caption{Centrality (upper panel), rapidity (middle panel), and transverse momentum (lower panel) dependence of $\Lambda$ polarization at $\sqrt{s_{\mathrm{NN}}}=2.55$ GeV in Ag+Ag collisions within the HADES experimental cuts. $4\times10^5$ events are used for each centrality. The experimental results are adopted from HADES \cite{HADES:2022enx}.}
    \label{fig:ag}
\end{figure}

\subsection{Differential analysis}

To better understand the differences between the simulation and data, we compare our simulation results with more differential data from the STAR and HADES experiments. Figure~\ref{fig:pt_y_cen} shows the centrality dependence (upper panel), rapidity dependence (middle panel) and $p_T$ dependence (lower panel) of the $\Lambda$ polarization in Au+Au collisions at $\sqrt{s_{\rm NN}}=3$~GeV. The trend of the centrality dependence is well reproduced with both our UrQMD simulations (CMF-EoS (blue) and HRG/CASCADE (orange)), while the HRG/CASCADE simulation gives a smaller polarization. Also, the rapidity dependence is similar for both EoS. In general, the model gives better results near mid-rapidity and becomes worse closer to the spectators. This is to be expected as the hyperons which decouple near the spectators cannot be considered to be in local equilibrium with the `medium' at freeze-out. The transverse momentum dependence in the models is essentially flat, while the data indicate a small downward trend at larger $p_T$. This may also be related to non-equilibrium effects, as high-momentum hyperons may deviate more from the local equilibrium assumption that is implied for the application of the thermal vorticity. In general we can say that both versions of the model are consistent with the data, within errors and show very similar systematic trends.  

Next we turn to a comparison with HADES data on the $\Lambda$ polarization in Ag+Ag collisions at $\sqrt{s_{\mathrm{NN}}}=2.55$ GeV. Figure \ref{fig:ag} shows the same quantities as before, this time for the model-data comparison of Ag+Ag collisions. In this case we can clearly see, that the model over-predicts the HADES data. This is especially the case at forward rapidities and large momenta. Again, that is the region where we expect non-equilibrium effects to play a stronger role which may indicate that at this lower beam energy and smaller system the assumption of full local spin-equilibrium is not valid anymore.  

\begin{figure}[!t]
    \centering
    \includegraphics[width=0.45\textwidth]{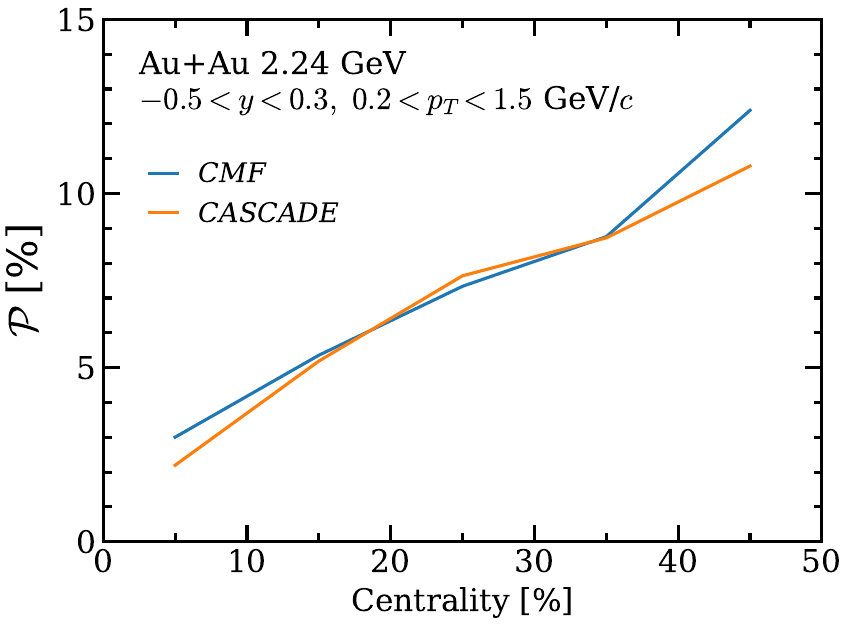}
    \includegraphics[width=0.45\textwidth]{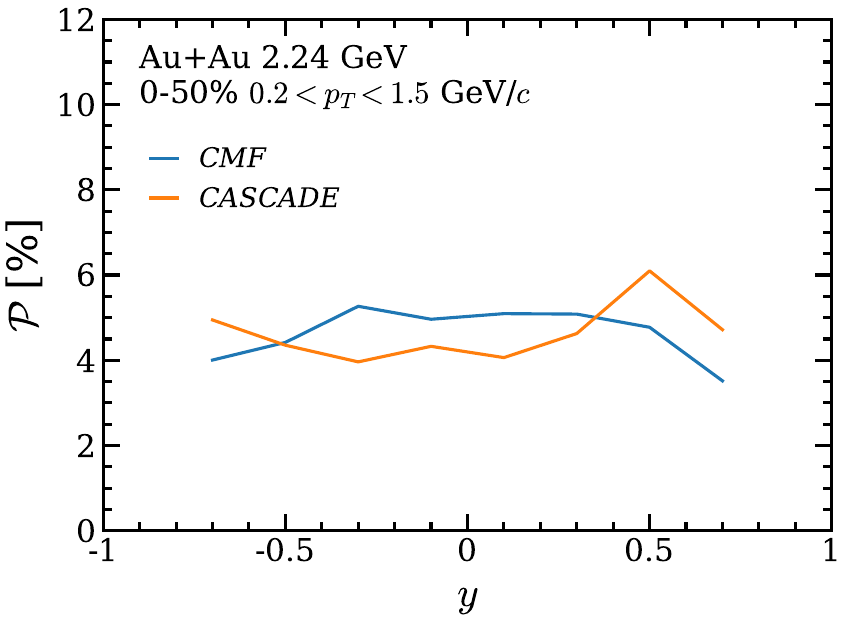}
    \includegraphics[width=0.45\textwidth]{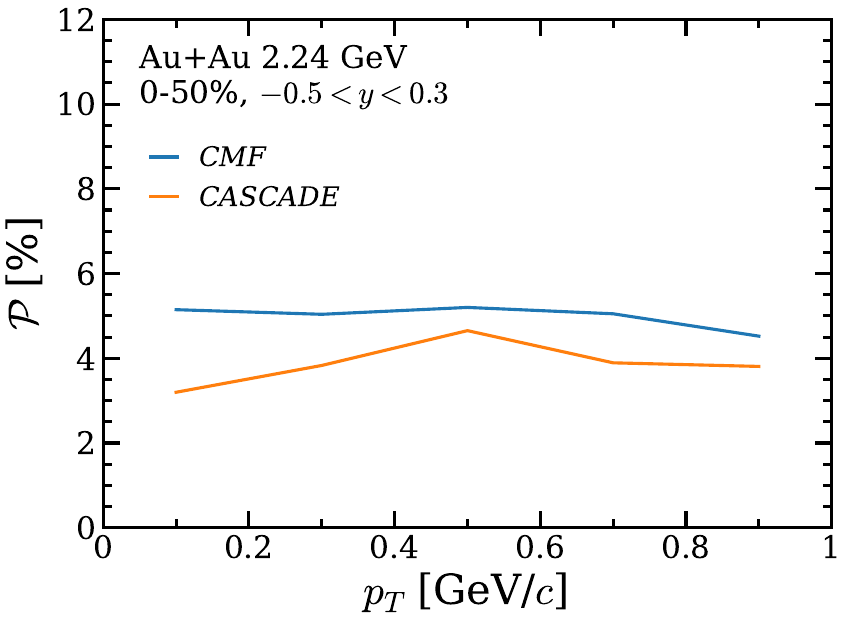}
    \caption{Predictions for the centrality (upper panel), rapidity (middle panel), and transverse momentum (lower panel) dependence of $\Lambda$ polarization in Au+Au collisions at ${E_{\mathrm{lab}}= 0.8 A}$~GeV. $2\times10^6$ events are used for each centrality.}
    \label{fig:2.24}
\end{figure}

As we have seen, the prediction based on a coarse grained thermal vorticity, has become less reliable for smaller systems and/or lower beam energies. It is therefore worthwhile to make predictions for upcoming HADES measurements of Au+Au at even lower beam energies i.e.~$\sqrt{s_{\mathrm {NN}}}= 2.24$ GeV, as the integrated polarization is still large in Fig.~\ref{fig:ener2050}. Figure \ref{fig:2.24} shows the centrality, rapidity, and $p_T$ dependence of the $\Lambda$ polarization in the current HADES acceptance. Again, the general trend observed at higher beam energies is the same. 

It should be noted that the theoretical curves shown in Figs.~\ref{fig:ag} and \ref{fig:2.24} are presented without uncertainty bands. The statistical uncertainty is expected to be approximately 5-10\% and become larger in the more peripheral centrality bins, at forward rapidities, and at high $p_\mathrm{T}$, due to the reduced $\Lambda$ hyperon yield in these regions. Nevertheless, the qualitative trends in the centrality, rapidity, and $p_\mathrm{T}$ dependence are expected to be more robust.
% than the absolute magnitude of the predicted polarization.
%The most notable effect is observed in the rapidity dependence, especially at rapidities close to the beam and target rapidity. 

\section{Discussion and outlook}
It was shown that the polarization of $\Lambda$ hyperons can be well described using UrQMD and coarse graining to calculate the thermal vorticity in heavy ion reactions at SIS/FAIR and RHIC-BES beam energies. Two equations of state for the simulation of the dynamical evolution of the systems were compared, the CMF-EoS (including potential interactions and a crossover transition at finite temperature) and the HRG/CASCADE mode (corresponding to a standard hadron resonance gas). Both EoS are consistent with available polarization data within the errors. The stiffer CMF EoS shows an increased polarization due to larger flow as compared to the softer HRG/CASCADE EoS. Comparing with data available from the HADES and STAR experiments it was observed that the description of the data worsens at larger rapidities and higher transverse momenta, a region where the assumption of local equilibration becomes questionable. Based on our model we have made predictions for $\Lambda$ polarization at even lower beam energies which showed that the polarization will likely not decrease as expected before. Our results indicate that the process leading to the large vorticity is driven primarily by the large shear in the baryon current created by its stopping. It is therefore to be expected, that the polarization can be related to the tilt of the emission source that has been observed in heavy ion reactions \cite{STAR:2026skb}. This relation will be studied in a forthcoming work in more detail.

While our study shows a clear EoS dependence of hyperon polarization, several effects still need further investigation. One is the loss/gain in polarization from feed-down to the $\Lambda$ hyperon, e.g. from the $\Sigma^0$ and $\Sigma^*(1385)$. In previous studies, the effect of feed-down was estimated as 10-20\%~\cite{Becattini:2016gvu,Xia:2019fjf,Becattini:2019ntv}.
Furthermore, the microscopic processes that transport spin are usually not included in transport models, which makes it difficult to incorporate non-equilibrium effects. Such effects are, however, important to explain the full differential dependencies of polarization and should be studied in more detail in future works.

\section*{Acknowledgments}
This work was supported in part by the National Key Research and Development Project of China under Grant No.~2024YFA1612500, No.~2022YFA1602303, No.~2022YFA1604900, the National Natural Science Foundation of China under Grants No. 12422509, No. 12375121, No.~12547102, No.~12025501, No.~12322508, and No.~124B2102, the STCSM under Grant No.~23590780100 and ~23JC1400200. 
D.-N.~Liu is also supported by the China Scholarship Council under Grant No.~202506100099. The computations in this research were performed using the CFFF platform of Fudan University.

 % No.~11891070, No.~11890714,
\bibliography{refs}

\end{document}